\documentclass[aps,prb]{revtex4}

\usepackage{color}
\usepackage{graphicx}
\begin{document}

\title{NMR properties of  the polar phase of superfluid $^3$He in anisotropic aerogel under rotation}

\author{V.P.Mineev}
\affiliation{Commissariat a l'Energie Atomique, INAC / SPSMS, 38054 Grenoble, France}

\begin{abstract}
The  polar phase of superfluid $^3$He is stable  in "nematically ordered"  densed aerogel. A rotating vessel with the polar superfluid
can be filled either by an array of the single quantum vortices
 or by an array of the half-quantum vortices.
 It is shown that the inhomogeneous distribution of the spin part of the order parameter arising in an array of half-quantum vortices in strong enough magnetic field tilted to the average direction of aerogel strands leads to the appearance of a satellite in the NMR signal shifted in the negative direction with respect to the Larmor frequency. The satellite is absent in the case of an array of single quantum vortices what allows to distinguish these two configurations.

The polar state in the anisotropic aerogel with lower density transforms at lower temperatures to the axipolar state.
The array of  half-quantum vortices  created 
in the polar phase
keeps its structure  under transition to the axipolar state. The temperature dependence of the vortex-satellite NMR  
frequency is found to be slower below the transition temperature  to the axipolar state.

\end{abstract}

\date{\today}
\maketitle

\section{ Introduction}
 About a decade ago it was predicted theoretically\cite{Aoyama} that liquid $^3$He in anisotropic aerogel transforms below the normal-superfluid phase transition  to the superfluid polar state. This was
 recently demonstrated experimentally\cite{Dmitriev}
using the dense anisotropic aerogel called  "nafen" consisting of  Al$_2$O$_3$ strands nearly parallel to one another and  having diameters 6-9 nm.
 The polar state is the superfluid equal spin pairing state with the order parameter given by the  product of the spin  ${\bf V}$ and the orbital ${\bf p}$ real unit vectors
\begin{equation}
A^{pol}_{\alpha i}=\Delta_{pol} V_\alpha p_ie^{i\phi}. 
\end{equation}
 The aerogel uniaxial anisotropy fixes the direction of the ${\bf p}$ vector along the $\hat z$ direction parallel to the aerogel strands.
 On the other hand,  the spin vector ${\bf V}$ must lie in the plane   perpendicular to the ${\bf p}$  direction what corresponds to the minimum of 
 the spin-orbital interaction  
 \begin{equation}
F_{so}=
\frac{2}{5}g^{pol}_d\left (({\bf V}{\bf p})^2-\frac{1}{3} \right ).
\end{equation}
Here, $g_d^{pol}\propto |\Delta_{pol}|^2$ is the amplitude of the dipole-dipole interaction  proportional to the square modulus of the polar state order parameter.

In what follows we consider the polar phase under rotation with an angular velocity parallel to the average direction of the aerogel strands.
In the equilibrium a rotating vessel with a  superfluid is filled by an array of parallel quantized vortices with the density $n_v=2\Omega/\Gamma$ such that the local average superfluid velocity is equal to the velocity of the normal component rotating with the angular velocity 
$\Omega$ of vessel. Here $\Gamma$ is the circulation quantum.  For the single quantum vortices  in the superfluid $^3$He it is equal  to  $\frac{h}{2m_3}=0.66\times 10^{-3}$ cm$^2$/s. 
Similar to  an Abrikosov lattice in the presence of inhomogeneities \cite{Larkin} 
  a long-range order in the arrangement of vortex lines is  spoiled   owing to the pinning of vortices  on the aerogel strands.

 An equilibrium lattice of vortices arises at slow enough finite $\Omega$ on cooling from the normal to the superfluid state. 
The  polar superfluid in rotating vessel can be filled \cite{Mineev2014} either by an array of the single quantum vortices with the density $n_v$, or by an array of the half-quantum vortices with the double density $2n_v$. 
 In the first case,  vortices do not disturb the homogeneous distribution of ${\bf V}$ vectors in the plane perpendicular to the aerogel strands. 
 In the second one, each
half-quantum vortex is accompanied by the basal plane $\pm\pi$  disclination in the  vector ${\bf V}$ field.  As in the rotating $^3$He-A in parallel plate geometry considered  by M.Salomaa and G.Volovik\cite{Volovik} the system of disclinations resembles a two-dimensional  electroneutral plasma such that the vortices with $+\pi$ and $-\pi$ disclinations alternate in the vortex lattice.
 The inhomogeneous distribution of vector ${\bf V}$ in the intervortex space
 around pair of half quantum vortices is shown in Fig.1.
 The half-quantum vortices accompanied by the disclinations in the  vector  ${\bf V}$ field are pinned to the aerogel strands. 
  
 A magnetic field $H>50G$   orients the ${\bf V}$ vector in the plane normal to the field direction.
 Hence, a constant magnetic field along the anisotropy axis does not disturb the vector ${\bf V}$ texture in rotating vessel.
 But a magnetic field directed at the angle
$\mu$ to the axis $\hat z$,  that is along  
\begin{equation}
 \hat{\bf h}=\hat y\sin\mu+\hat z\cos\mu,
 \end{equation}
changes  the vector  ${\bf V}$ distribution. 
Namely, when the basal plane field component $H\sin\mu$ exceeds the spin-orbit interaction, the vector ${\bf V}$ texture  consists: (i) from the homogeneous  distribution ${\bf V}\parallel \hat x$
in almost all inter-vortex space what corresponds to minimum both the magnetic and the spin-orbit energy
 and (ii) from the narrow walls or planar solitons  with thickness  of the order of dipole length  $\xi_d\sim 10^{-3}$ cm  connecting the opposite sign $\pm\pi$ disclinations inside of which  vector ${\bf V}\perp{\bf H}$ performs rotation on the angle $\pi$ around field direction.

 \begin{figure}
\includegraphics[height=.5\textheight]{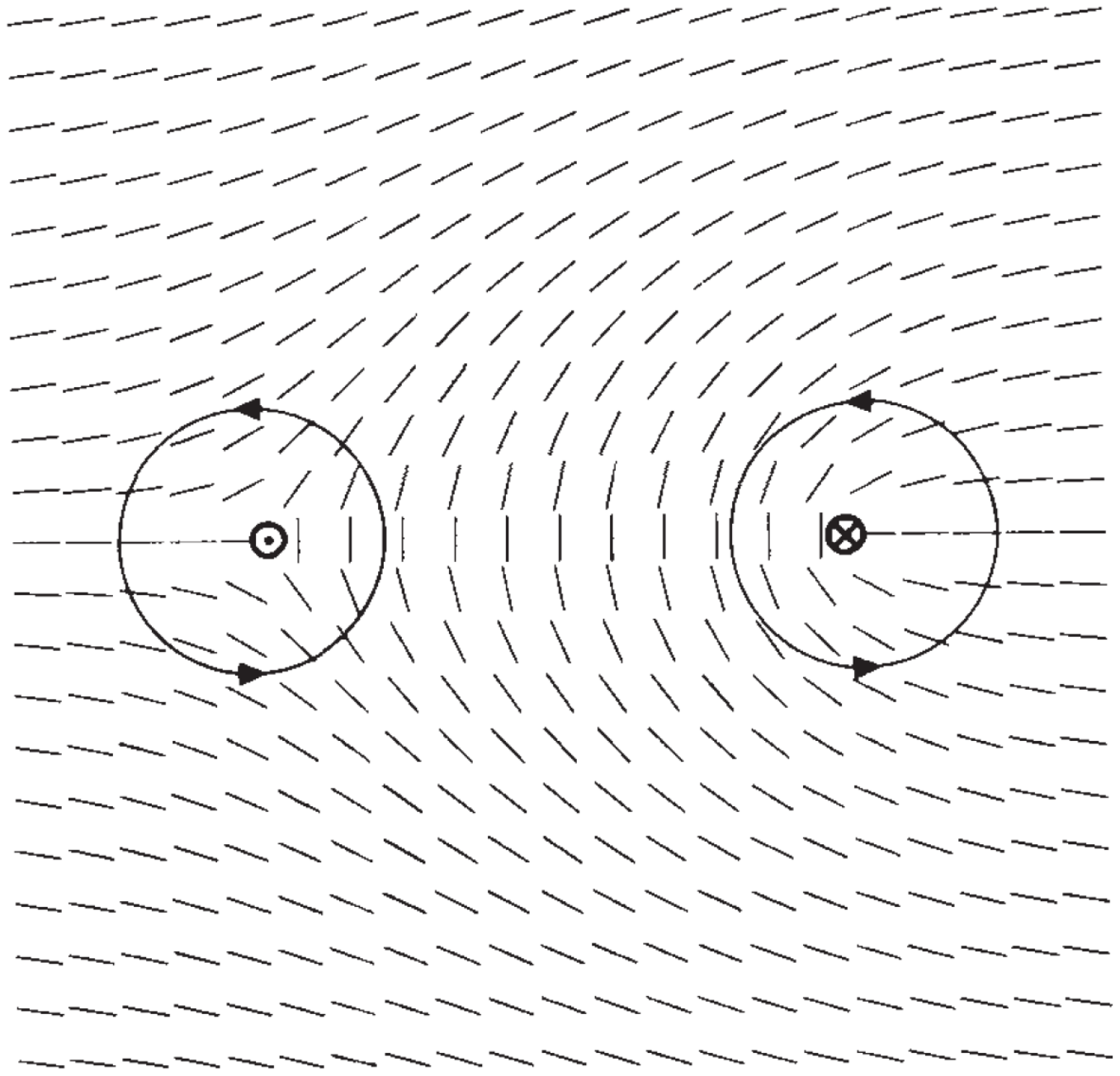}
\caption{Distribution of ${\bf V}$ vector in the  $(x,y)$ plane 
  around pair of half-quantum vortices.   Directions of the ${\bf V}$ vector shown by traits display pair of  disclinations with opposite Frank indices: on circling each vortex line the phase of the order parameter 
changes by $\pi$ while
${\bf V}$ rotates by $+\pi$ and by $-\pi$, resulting in single-valuedness of the order parameter Eq. (1).}
\end{figure}

The continuous NMR signal  can be roughly represented as originating from the homogeneous region of vector ${\bf V}$ texture and from the solitons connecting disclinations. As we shall see the NMR frequencies from the homogeneous region and from the solitons are different. Hence, they reveal themselves as the  two separated peaks in the NMR signal.
In the case of the single quantum vortices the vector  ${\bf V}$  texture is homogeneous, the soliton peak
is absent  that allows to distinguish these two configurations.

In what follows we consider the Nuclear Magnetic Resonance properties   in the rotating polar  phase under magnetic field and their modification below the phase transition to the so-called axipolar phase what was revealed in an anisotropic aerogel of lower density \cite{Dmitriev}.

This paper is based on the approach developed by the author in Ref.4.  The more rigorous derivation and the corresponding experimental results are reported in the preprint \cite{Autti}.

\section{Nuclear Magnetic resonance Properties}

The shift of the NMR frequency from the Larmor frequency is determined by the spin-orbital energy which in the pure polar state has the following form\cite{Mineev2014}
 \begin{equation}
F_{so}=2(C_1+C_2)\Delta_{pol}^2\langle ({\bf V}(t)\hat z)^2\rangle.
\label{sopol}
\end{equation}
Here,  $C_1$ and  $C_2$ are the constants of  the dipole-dipole interaction in an anisotropic aerogel, $\Delta_{pol}$ is the amplitude of the order parameter of the polar state, the angular brackets indicate the time averaging over fast precessional motion of
the ${\bf V}(t)$ vector around the direction of the precessing with  resonance  frequency magnetization\cite{Fomin,Gongadze}.
Taking into account, that (i) the equilibrium vector ${\bf V}$ distribution inside the solitons undergoes rotation on the angle $\varphi$ varying  the interval $(0,\pi)$  around the field direction, (ii)
the external magnetic field is directed at the angle $\mu$ to the anisotropy axis $\hat z$  in the $(y,z)$ plane, and  (iii) the  precessing magnetization is tilted at the angle $\beta$ to the field direction, we have
\begin{equation}
{\bf V}(t)=R_x(\mu)R_z(-\omega_Lt)R_y(\beta)R_z(\omega_Lt+\varphi)\hat x,
\end{equation}
where $R_x(\mu)$ is the matrix of rotation  around the $x$ axis on  angle $\mu$ with the other matrices similarly defined.
Performing the time averaging  we obtain  the angular dependence of the dipole energy
\begin{equation}
\langle ({\bf V}(t)\hat z)^2\rangle=\frac{1}{4}\sin^2\mu\left[(\cos\beta+1)^2\sin^2\varphi+\frac{1}{2}(\cos\beta-1)^2\right]+\frac{1}{2}\cos^2\mu(1-\cos^2\beta).
\label{6}
\end{equation}
The frequency  shift  of the transverse NMR from the Larmor value is given by  
\begin{equation}
2\omega_L\Delta\omega=-\frac{2\gamma^2}{\chi}\frac{\partial~\langle F_{so}\rangle}{\partial \cos\beta}=-\Omega_{pol}^2\frac{\partial~\langle ({\bf V}(t)\hat z)^2\rangle}{\partial \cos\beta},
\label{shift}
\end{equation}
\begin{equation}
\Omega_{pol}^2=\frac{2\gamma^2}{\chi}2(C_1+C_2)\Delta_{pol}^2.
\label{ampl}
\end{equation}
Here $\gamma$ is the $^3$He nuclear  gyromagnetic ratio, and $\chi$ is the paramagnetic susceptibility.
Substituting Eq.(\ref{6}) in Eq. (\ref{shift}) we obtain
\begin{equation}
2\omega_L\Delta\omega=\Omega_{pol}^2\left [\cos\beta+\frac{1}{4}\sin^2\mu(1-5\cos\beta) -\frac{1}{2}\sin^2\mu\sin^2\varphi(1+\cos\beta)\right].
\end{equation}
In  regions of  homogeneous ${\bf V}$ vector  $\varphi=0$, thus
\begin{equation}
2\omega_L\Delta\omega=\Omega_{pol}^2\left [\cos\beta+\frac{1}{4}\sin^2\mu(1-5\cos\beta)
\right].
\end{equation}

In the case of continuous NMR $\beta=0$ and 
\begin{equation}
2\omega_L\Delta\omega=\Omega_{pol}^2\cos^2\mu
\end{equation}
such that with the external field lying in the basal plane ($\mu=\pi/2$)  the resonance frequency coincides with the Larmor frequency.
Thus, in the case of singly-quantized vortices the NMR shift is equal to zero.

In  solitons regions the vector ${\bf V}$ distribution is inhomogeneous, and angle $\varphi$  varying in the interval $(0,\pi)$. 
Then, for the continuous NMR $\beta=0$ 
we obtain
\begin{equation}
2\omega_L\Delta\omega=\Omega_{pol}^2(\cos^2\mu-\sin^2\mu\sin^2\varphi).
\end{equation}
Thus for a basal plane field orientation we obtain the satellite frequency shifted in respect of the  Larmor frequency
 \begin{equation}
2\omega_L\Delta\omega=-\Omega_{pol}^2\sin^2\varphi.
\label{satfreq}
\end{equation}
This frequency shift obtained in the paper \cite{Mineev2014} has been averaged  over the $\varphi$ distribution
giving $2\omega_L\Delta\omega=-\Omega_{pol}^2/2$.
Actually, the inhomogeneous angle $\varphi$ distribution inside the planar solitons with thickness $\xi_d$ creates an additional torque
due to the increase  of gradient energy \cite{footnote}. As a result the satellite frequency shift acquires the following form \cite{Maki}
 \begin{equation}
2\omega_L\Delta\omega=-\Omega_{pol}^2\left [\sin^2\varphi+\xi_d^2(\nabla \varphi)^2\right ].
\label{sate}
\end{equation}
Averaging this expression over the angle  $\varphi$ distribution inside the soliton yields  the frequency of the satellite  line in the continuous NMR signal in  transverse field
 \begin{equation}
2\omega_L\Delta\omega=-\Omega_{pol}^2.
\label{satel}
\end{equation}
 The frequency $\Omega_{pol}$ expressed by Eq. (\ref{ampl}) through the amplitude
  of the polar state order parameter. 
  The latter  near the phase transition from the normal state is
\begin{equation}
\Delta_{pol}^2=\frac{\alpha_0(T_{c1}-T)}{2\beta_{12345}},
\end{equation}
hence the satellite frequency shift is written as
\begin{equation}
2\omega_L\Delta\omega=-\frac{2\gamma^2}{\chi}2(C_1+C_2)\frac{\alpha_0(T_{c1}-T)}{2\beta_{12345}},
\end{equation}
where 
$$\beta_{12345}=\beta_1+\beta_2+\beta_3+\beta_4+\beta_5$$
is the sum of coefficients in the fourth order terms in the Ginzburg-Landau expansion of the free energy
\begin{eqnarray}
&F_{cond}=\alpha A_{\alpha i}^\star A_{\alpha i}+\eta_{ij}A_{\alpha i}A^\star_{\alpha j}\nonumber\\
&+\beta_1|A_{\alpha i}A_{\alpha i}|^2+\beta_2A_{\alpha i}^\star A_{\alpha j}A_{\beta i}^\star A_{\beta j}+
\beta_3A_{\alpha i}^\star A_{\beta i}A_{\alpha j}^\star A_{\beta j}+\beta_4(A_{\alpha i}^\star A_{\alpha i})^2+\beta_5A_{\alpha i}^\star A_{\beta i}A_{\beta j}^\star A_{\alpha j},
\label{GL}
\end{eqnarray}
$$
\alpha=\alpha_0(T-T_{c}),~~~~~ \eta_{ij}=\eta\left (\begin{array}{ccc}1&0&0\\
 0&1&0\\
 0&0&-2\end{array}\right ), 
 $$ 
 $T_c=T_c({P})$ is the transition temperature in a superfluid state suppressed   in respect of transition temperature in the bulk liquid  $T_{c0}({P})$ due to the isotropic part of quasiparticles scattering on the aerogel strands and 
\begin{equation}
T_{c1}=T_c+2\frac{\eta}{\alpha_0}
\label{1}
\end{equation}
is the transition temperature from the normal to the polar state.

\section{Axipolar state}

There are two types of  anisotropic aerogel samples where the transition into the pure polar phase occurs.\cite{Dmitriev} They consist of Al$_2$O$_3$ strands and have porosities of 97.8 percent (sample"nafen-90" with overall density 90 mg/cm$^3$) and 93.9 percent ("nafen-243" with density 243 mg/cm$^3$).
In the dense anisotropic aerogel (nafen-243) the superfluid polar state arising below 
the critical temperature of phase transition from the normal state exists up to the lowest temperatures  reachable 
experimentally \cite{Dmitriev}.  On the other hand, in the anisotropic aerogel of lower density (nafen-90) \cite{Dmitriev} as temperature decreases the polar phase  gives way to the more energetically favorable axipolar phase. The latter also has the order parameter given by the product of the spin and the orbital vectors\cite{Mineev2014}
\begin{equation}
A^{axipolar}_{\alpha i}=V_\alpha \left[a\hat z_i+ib(\hat x_i\cos\theta({\bf r})+\hat y_i\sin\theta({\bf r}))\right]e^{i\phi},
\label{op}
\end{equation} 
where the angle $\theta({\bf r})$ determines the local direction of the Cooper pair angular momentum $\hat l=-\hat x\sin\theta({\bf r})+\hat y\cos\theta({\bf r})$ randomly distributed in the basal plane according to Imry-Ma ideology.
Hence, the axipolar state
can support  the existence of the half-quantum vortices.
These vortices,  if they already formed in the polar state, are preserved under
the second-order phase transition from the polar to the axipolar state. Here we  find the modification of the satellite NMR frequency below the phase transition  from the polar to the axipolar state.
  
The axipolar phase arises
below the critical temperature\cite{Mineev2014}
\begin{equation}
T_{c2}=T_c-\frac{\eta}{\alpha_0}\frac{3\beta_{345}-\beta_{12}}{2\beta_{12}},
\label{c2}
\end{equation}
where 
$$
\beta_{12}=\beta_1+\beta_2,~~~~~\beta_{345}=\beta_3+\beta_4+\beta_5.
$$
are the sums of coefficients in the fourth order terms in the Ginzburg-Landau expansion of the free energy Eq.(\ref{GL}).

Similar to the polar state the minimum of spin-orbital energy in the axipolar state is carried out when the ${\bf V}$ vector is perpendicular to the anisotropy axis $\hat z$. Hence,  the strong enough magnetic field tilted to the strands  forms the homogeneous  regions,  where the vector ${\bf V}$ lies in the direction perpendicular both to the arerogel strands and to the field direction, divided  by the narrow  domain walls - planar solitons,  connecting the half-quantum vortices with the opposite sign disclinations, inside of which the vector ${\bf V}\perp{\bf H}$ performs rotation on the angle $\pi$ around the field direction. 
The frequency of the satellite in the NMR signal  in the transverse field corresponding to the inhomogeneous   order parameter  distribution inside the planar solitons 
is described by the formula  similar to Eq. (\ref{satel}) for the polar phase
\begin{equation}
2\omega_L\Delta\omega
\approx-\Omega_{axipol}^2
\label{sat2}
\end{equation} 
 but with the different amplitude 
\begin{equation}
\Omega^2_{axipol}=\frac{2\gamma^2}{\chi}2(C_1+C_2)\frac{\alpha_0}{2\beta_{12345}}\left [T_{c1}-T +r(T-T_{c2})\right  ]
\label{r}
\end{equation}
found in the Ref.4. Here
\begin{equation}
1-r=1-\frac{1}{4\beta_{345}}\left [2(\beta_{345}-\beta_{12})+\frac{C_1}{C_1+C_2}\beta_{12345}   \right ]
\end{equation}
is the ratio of the slopes of the temperature dependence of the phase shift below and above $T_{c2}$.
We note that in Ref.4 there was  a misprint in the expression  for $\Omega^2_{axipol}$ which is corrected here.

To estimate $r$  we neglect by the anisotropic part of the spin-orbital energy by setting $C_2=0$, and using the values of  the $\beta_i$ coefficients in the weak coupling approximation \cite{Mineev2014}.
We obtain
\begin{equation}
r=\frac{3\beta_{345}-\beta_{12}}{4\beta_{345}}=\frac{5}{8}=0.625.
\end{equation}
Thus, the slope of the $\Delta\omega$ temperature dependence below $T_{c2}$ begins to decrease.

\section{Conclusion}

Liquid $^3$He filling strongly anisotropic dense aerogel  transforms below the critical temperature to the superfluid polar phase\cite{Dmitriev}.
Slow enough liquid cooling through the phase transition  in a rotating vessel with  angular velocity parallel to the aerogel strands can create an equilibrium density of  quantum vortices  either with single or with half quantum of circulation forming  rotation of the superfluid component with average velocity equal to  the velocity of the normal component. Each half-quantum vortex is accompanied by a $\pm\pi$  disclination in the field of the 
${\bf V}$ vector  that is in  the spin part of the order parameter. In a constant magnetic field with magnitude  stronger than the dipole-dipole forces 
and  tilted to the aerogel strands almost all intervortex space is filled by a homogeneous distribution of vector ${\bf V}$ lying in the direction perpendicular both to the arerogel strands and the field direction. At the same time all pairs of disclinations of the opposite sign are connected by the narrow walls of an inhomogeneous ${\bf V}$ vector distribution with thickness $\xi_d$.

We have shown that the walls regions create  the satellite in the continuous NMR signal  with frequency smaller than the Larmor frequency.
The single quantum vortices do not create an inhomogeneity of the spin part of the order parameter. This case  satellite
is absent  what allows to distinguish single-quantum and half-quantum vortices configurations.

In the anisotropic aerogel of the lower density \cite{Dmitriev} as temperature decreases the polar phase  transforms  to the more energetically favorable axipolar phase. The latter  can also support  the existence of half-quantum vortices pinned by the aerogel strands.
Below the phase transition to the axipolar state
the temperature dependence of the satellite NMR frequency corresponding to intervortex stripes  is reduced.

\section*{Acknowledgements}
I am indebted to V. Eltsov and V. Dmitriev
for useful discussions and especially to G. Volovik  for numerous and careful comments which have allowed me to correct the initially not properly written text.


\begin{thebibliography}{20}
\bibitem{footnote} I am indebted to G.Volovik who has pointed me on the additional gradient energy term derived in the paper \cite{Maki}.
\end{thebibliography}
\end{document}